\documentclass[aps,prl,reprint,superscriptaddress]{revtex4-2}
\usepackage{blindtext}
\usepackage{xcolor}
\usepackage{hyperref}
\usepackage{epsfig}
\usepackage{subfigure}
\usepackage{graphicx}
\usepackage[toc,page]{appendix}
\usepackage{epstopdf}
\usepackage{amsmath,amssymb,amsfonts}
\usepackage{array}
\usepackage{url}
\usepackage{hyperref}
\usepackage{lineno}
\usepackage{xspace}
\usepackage{listings}
\usepackage{float}
\usepackage{subfigure}
\usepackage{subfiles}
\usepackage{placeins}

\begin{document}
\title{Measurement of Fifth- and Sixth-Order Fluctuations of (Net-)proton Number in Au+Au Collisions from Phase II of the Beam Energy Scan Program at RHIC}
\author{The STAR Collaboration}
\date{\today}

  \begin{abstract}
We report high-statistics measurements of fifth- and sixth-order factorial cumulants and cumulant ratios of (net-)proton multiplicity distributions in Au+Au collisions at $\sqrt{s_{NN}} = 7.7$--27\,GeV, using data from the STAR experiment collected during the Beam Energy Scan Phase~II at RHIC. Protons and antiprotons are identified at midrapidity ($|y| < 0.5$) with transverse momentum $0.4 < p_T < 2.0$\,GeV/$c$. The proton factorial cumulants $\kappa_4$, $\kappa_5$, and $\kappa_6$ increase with order but exhibit no sign alternation within current uncertainties, offering no evidence for a two-component structure in the proton multiplicity distribution, as might be expected near a first-order phase transition. The cumulant ratios \( C_{5}/C_{1} \) and \( C_{6}/C_{2} \) fluctuate around zero in collisions at 0--40\% centrality. The results are consistent with both the negative predictions from lattice QCD (LQCD) and the positive trends obtained from the Ultra-relativistic Quantum Molecular Dynamics (UrQMD) model. At $\sqrt{s_{NN}} \gtrsim 27$\,GeV, the $C_4/C_2$ and $C_5/C_1$ results are compatible with predictions from lattice QCD, functional renormalization group (FRG), and hadron resonance gas (HRG) models, while UrQMD describes the data better at lower energies. These measurements place constraints on baryon number fluctuations and offer valuable insights into the QCD phase structure.
\end{abstract}


\maketitle


\section{Introduction}

Fluctuations of conserved quantities, such as baryon number, electric charge, and strangeness, in relativistic heavy-ion collisions offer a powerful tool to probe the phase structure of Quantum Chromodynamics (QCD)~\cite{Stephanov:1998dy,Stephanov:2004wx,Asakawa:2000wh,Gupta:2011wh,Pandav:2022xxx}. In particular, higher-order cumulants of net-proton multiplicity distributions are of significant interest due to their sensitivity to the correlation length and their potential to reveal signatures of the QCD critical point~\cite{Athanasiou:2010kw,Stephanov:2011pb}.

The Beam Energy Scan (BES) program at the Relativistic Heavy Ion Collider (RHIC) was designed to explore the QCD phase diagram by varying the collision energy of heavy nuclei, thereby scanning a broad range of temperatures and baryon chemical potentials ($\mu_B$)~\cite{STAR:2010vob,STAR:2017sal}. Cumulants ($C_n$) of the net-particle number ($N$) distribution serve as key observables in this study. 
The fifth- and sixth-order cumulants used here are defined by 
$C_{5} = \langle (\delta N)^{5} \rangle - 10\,\langle (\delta N)^{3} \rangle \langle (\delta N)^{2} \rangle$ 
and 
$C_{6} = \langle (\delta N)^{6} \rangle - 15\,\langle (\delta N)^{4} \rangle \langle (\delta N)^{2} \rangle - 10\,\langle (\delta N)^{3} \rangle^{2} + 30\,\langle (\delta N)^{2} \rangle^{3}$, 
with $\delta N = N - \langle N \rangle$. Based on Phase I of the BES program (BES-I), the STAR collaboration reported net-proton cumulants up to fourth order ($C_1-C_4$)~\cite{STAR:2013gus,STAR:2020tga}, later extending to fifth- and sixth-order cumulants ($C_5$ and $C_6$) with limited statistics~\cite{STAR:2021rls,STAR:2022vlo}. These early results suggested a non-monotonic energy dependence, but were subject to large statistical and systematic uncertainties, particularly at lower beam energies. These indications highlighted the need for the substantially increased statistics and improved detector coverage achieved in BES-II.

Phase II of the BES program (BES-II) introduced major detector upgrades, including the inner Time Projection Chamber (iTPC)~\cite{ref_iTPC_BESII}, the Event Plane Detector (EPD)~\cite{Adams:2019fpo}, and the end-cap Time-of-Flight (eTOF)~\cite{STAR:2016gpu}, which improved particle identification and extended the acceptance~\cite{ref_BESII_white}. Along with enhanced beam luminosity and longer run times, these improvements enabled STAR to collect high-statistics Au+Au collision data at $\sqrt{s_{NN}} = 7.7$--27~GeV. For example, at $\sqrt{s_{NN}} = 19.6$~GeV, an 18-fold increase~\cite{STAR:2025zdq} in statistics was achieved compared to the earlier BES-I~\cite{STAR:2021iop}. This dataset spans a baryon chemical potential ($\mu_B$) range of approximately 400--150~MeV~\cite{STAR:2017sal}, thereby enhancing the sensitivity of measured higher-order cumulants at the lower collider energies.

Recent advances in QCD thermodynamics offer theoretical benchmarks for the behavior of higher-order cumulants near the critical region. A negative sign of $C_5/C_1$ and $C_6/C_2$ is predicted by lattice QCD and functional methods that incorporate the quark–hadron crossover transition~\cite{Bazavov:2017dus,Fu:2021oaw}. Furthermore, lattice QCD predicts a specific ordering of cumulant ratios in the crossover regime, $C_3/C_1 > C_4/C_2 > C_5/C_1 > C_6/C_2$~\cite{Bazavov:2017dus}. In addition, factorial cumulants ($\kappa_n$)~\cite{Bzdak:2016sxg} have been proposed as useful observables for studying first-order phase transitions. They can be expressed in terms of the ordinary cumulants ($C_n$). 
The fifth- and sixth-order factorial cumulants used in this work are given by:   
$\kappa_{5} = 24C_{1} - 50C_{2} + 35C_{3} - 10C_{4} + C_{5}$  
and  
$\kappa_{6} = -120C_{1} + 274C_{2} - 225C_{3} + 85C_{4} - 15C_{5} + C_{6}$.They are expected to grow in magnitude with order, and a sign-alternating pattern is predicted near a first-order phase transition or in the presence of a two-component structure~\cite{Bzdak:2018axe,Bzdak:2018uhv}. Thus, the presence—or absence—of these features serves as a key diagnostic of the QCD phase transition and the proximity to a critical point.


\begin{figure*}[!htbp]
\centering
  \includegraphics[scale=0.69]{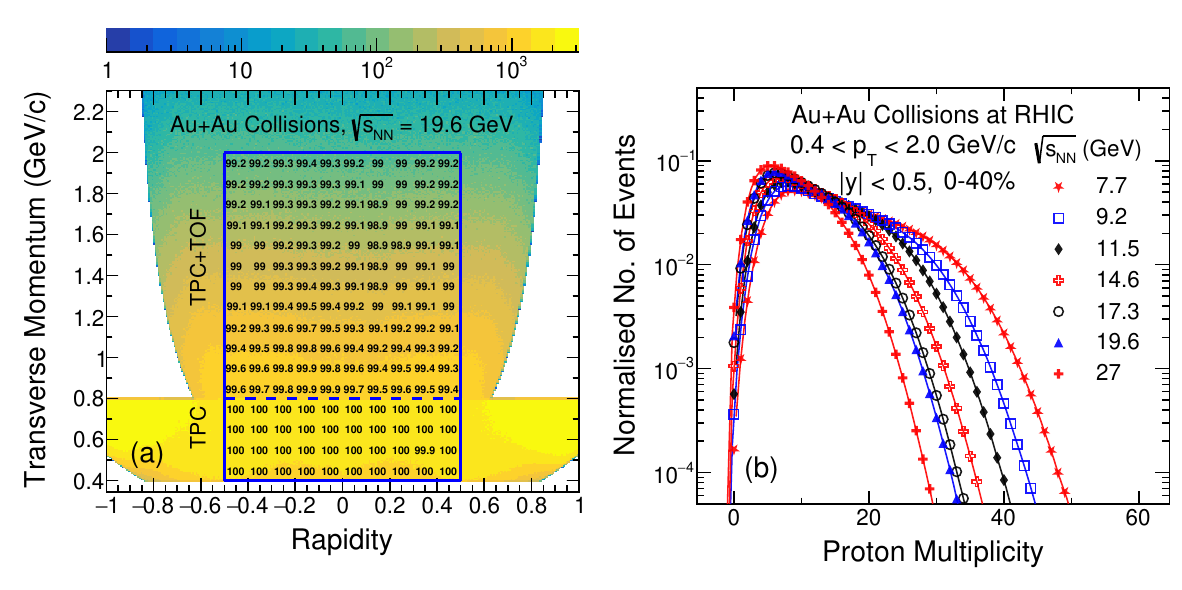}
 \caption{(a) Proton acceptance in rapidity ($y$) vs.\ transverse momentum ($p_{\mathrm T}$) for Au+Au collisions at $\sqrt{s_{NN}} = 19.6$~GeV from BES-II. The blue box indicates the kinematic window used in this analysis. The TPC is used for PID in the range $0.4 < p_T < 0.8$~GeV/$c$, while both the TPC and TOF are used for $0.8 < p_T < 2.0$~GeV/$c$. The numbers within each ($y$, $p_T$) bin represent the purity of proton identification. (b) Event-by-event proton multiplicity distributions for 0--40\% centrality of Au+Au collisions at BES-II energies ($\sqrt{s_{NN}} = 7.7$--27~GeV).}
  \label{figure1}
\end{figure*}

Recently, the STAR collaboration reported high-precision measurements of net-proton cumulants up to fourth order from BES-II~\cite{STAR:2025zdq}, covering Au+Au collisions over the same energy range. Relative to expectations from models without a critical point, and in contrast to STAR data for peripheral collisions, the most central \( C_{4}/C_{2} \) is seen to exhibit a minimum in collision energy dependence at \(\sqrt{s_{NN}} = 19.6\)~GeV with a significance of about 2--5\(\sigma\). These measurements, with improved precision and systematic control, offer strong constraints on the QCD phase structure and provide a crucial reference for the higher-order cumulant measurements presented here.

In this work, we present high-statistics measurements of the fifth- and sixth-order factorial cumulants and cumulant ratios of (net-)proton multiplicity distributions in Au+Au collisions, based on BES-II data collected by STAR. These results allow direct comparisons with theoretical predictions from  lattice QCD~\cite{Bazavov:2017dus,Bazavov:2020bjn}, FRG calculations~\cite{Fu:2021oaw}, and hadronic transport models such as UrQMD~\cite{Bass:1998ca}.

\section{Experiment and Data Analysis}

\subsection{Dataset and Event Selection}

This analysis is based on Au+Au collision data at $\sqrt{s_{NN}} = 7.7$--27~GeV, recorded by the STAR detector during BES-II at RHIC. A minimum-bias trigger~\cite{Adler:2000bd,Llope:2003ti} is used to select Au+Au collision events that generate signals in the trigger detectors exceeding the noise threshold. Events are required to have a primary vertex located within $\pm$50~cm along the beam axis (reduced to $\pm$27~cm at $\sqrt{s_{NN}} = 27$~GeV, where the iTPC upgrade was unavailable) and within 1~cm radially from the TPC center.

Beam-related backgrounds, pile-up, and poorly timed events are removed using standard STAR data-quality cuts~\cite{STAR:2021iop,STAR:2025zdq}. The number of good-quality events ranges from 45 million at $\sqrt{s_{NN}} = 7.7$~GeV to 270 million at 19.6~GeV~\cite{STAR:2025zdq}. Quality assurance (QA) procedures are implemented at the run, event, and track levels to ensure data integrity for the higher-order cumulant analysis. Run-level QA monitors key observables such as charged-particle multiplicity, TOF-matched track counts, and vertex distributions to reject anomalous runs. Event-level QA is applied to enforce vertex selection criteria and to remove pileup and other poor-quality events arising from unstable beam conditions or imperfect space-charge calibration in the TPC. Track selection requires at least 20 TPC space points, a distance of closest approach (DCA) to the primary vertex of less than 1 cm, and kinematic acceptance within $|y| < 0.5$ and $0.4 < p_T < 2.0$~GeV/$c$.

Particle identification (PID) for protons and antiprotons is carried out using ionization energy loss ($dE/dx$) in the TPC, supplemented with timing information from the TOF detector. The PID selection employs variable $n\sigma$~\cite{STAR:2017tfy}, defined as the deviation (in units of standard deviations) between the measured TPC $dE/dx$ and the expected values for protons~\cite{BICHSEL2006154}. These selection criteria are optimized to maintain stability across collision centralities and data-taking periods. As an example, the kinematic acceptance for proton selection at $\sqrt{s_{NN}} = 19.6$~GeV is shown in Fig.~\ref{figure1}(a). These criteria yield a high proton purity of approximately 99\% or greater at all the energies analyzed, indicating high-quality particle identification essential for the fluctuation analysis.

Collision centrality is determined using the multiplicity estimators RefMult3 and RefMult3X~\cite{STAR:2025zdq}, based on charged-particle counts within $|\eta|<1$ and $|\eta|<1.6$, respectively, with identified protons and antiprotons excluded to suppress self-correlations. RefMult3 follows the definition used in BES-I, while RefMult3X exploits the extended iTPC acceptance to $|\eta|<1.6$, thereby incorporating more tracks. The increased multiplicity range enhances centrality resolution, making RefMult3X the more precise estimator~\cite{STAR:2025zdq}.

These comprehensive selection and QA procedures ensure a high-purity, well-characterized dataset suitable for precision fluctuation analyses and reliable interpretation of higher-order cumulant observables.


\subsection{Corrections}

Raw event-by-event (net-)proton multiplicity distributions, shown in Fig.~\ref{figure1}(b) for the 0--40\% centrality range, serve as the basis for fluctuation analyses. To extract meaningful physics, the measured cumulants and factorial cumulants are corrected for finite detection efficiency using the binomial response method, which assumes independent particle detection~\cite{Luo:2018ofd,Nonaka:2017kko,Bzdak:2012ab}. Proton and antiproton detection efficiencies are determined by embedding Monte Carlo tracks, propagated through GEANT~\cite{Brun:1994aa} to model the STAR detector response, into real events. The efficiencies are evaluated as functions of transverse momentum ($p_T$), rapidity, centrality, and vertex position. Acceptance effects are controlled by restricting the analysis to a fiducial phase space of $|y| < 0.5$ and $0.4 < p_T < 2.0$~GeV/$c$, where tracking and PID performance are stable and well understood. To account for volume fluctuations within wide centrality bins, a centrality bin width correction (CBWC) is applied~\cite{Luo:2013bmi}, wherein cumulants are computed in fine multiplicity intervals and averaged over the wider centrality bin. 

These corrections are essential for isolating genuine fluctuation signals. They also simplify the interpretation of comparisons between data and theoretical models.

\subsection{Uncertainties}

Statistical uncertainties on cumulants and factorial cumulants are estimated using the bootstrap resampling technique~\cite{Efron:1979bxm,Luo:2014rea,Pandav:2018bdx}. Each bootstrap sample is constructed by randomly selecting, with replacement, the same total number of events as in the original dataset. A total of 50 such bootstrap samples are generated, and cumulants are calculated for each one. The statistical uncertainty is taken as the standard deviation of the cumulant values obtained from these bootstrap samples. This model-independent approach effectively captures non-Gaussian features and bin-to-bin correlations and is particularly suited to higher-order cumulants, where conventional error propagation may be insufficient~\cite{Pandav:2018bdx}.

Systematic uncertainties are evaluated by varying key analysis parameters, including PID selection criteria, track quality criteria, background estimates, and the track reconstruction efficiency factor. These variations are taken to be the same as in BES-I~\cite{STAR:2021iop}, except for the reconstruction efficiency factor, which is now varied by 2\% instead of 5\%, reflecting the improved tracking performance after the iTPC upgrade. For each variation, cumulants are 
recalculated, and the Barlow criterion~\cite{Barlow:2002yb} is applied as follows:

\begin{equation}
\Delta > \sqrt{|\sigma_{\text{stat.,def}}^2 - \sigma_{\text{stat.,var}}^2|}
\label{eq:stat_diff}
\end{equation}

Here, $\Delta$ denotes the absolute difference between the values obtained with the default selection criteria and those from the systematic variation. The terms $\sigma_{\text{stat.,def}}$ and $\sigma_{\text{stat.,var}}$ represent the statistical uncertainties of the default and varied results, respectively.
Only statistically significant deviations (those satisfying the above equation) are included in the systematic uncertainty. Assuming the different sources are uncorrelated, the total systematic error is computed by summing in quadrature, for each parameter tested, the RMS of the spread in statistically significant deviations. The DCA and PID variations are the two dominant sources of systematic uncertainty across the analyzed collision energies. For example, for the net-proton $C_6/C_2$ in 0-40\% centrality class at $\sqrt{s_{NN}}=19.6$ GeV, the DCA and PID contributions account for 93\% and 32\% of the total systematic uncertainty (\(\sigma_{\text{sys}}\)), respectively.

\section{Results}

\begin{figure*}[!htbp]
\centering
  \includegraphics[scale=0.98]{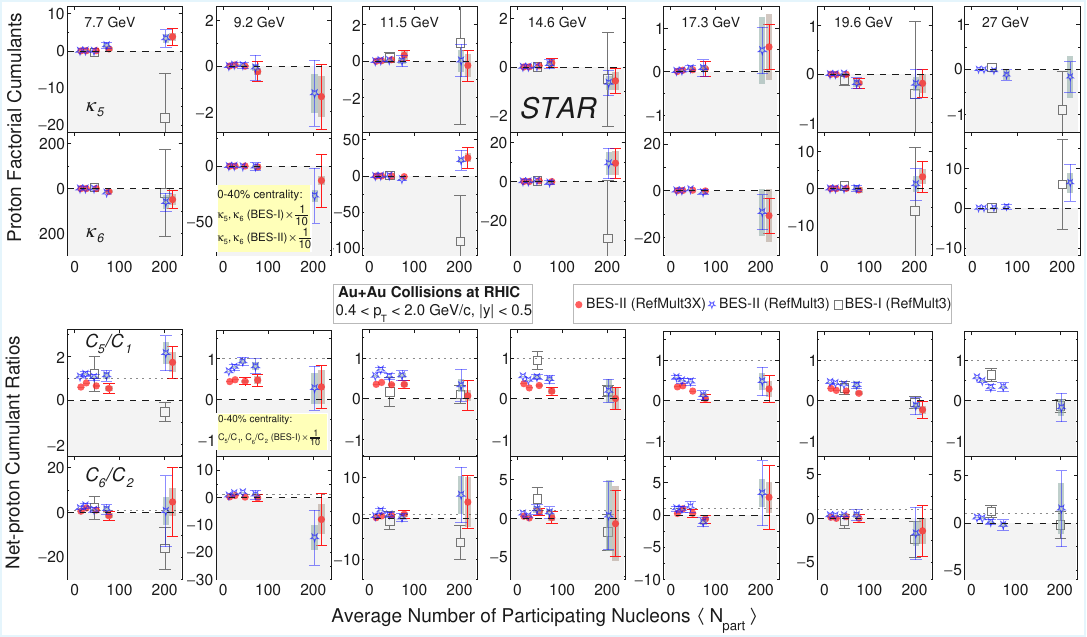}
  \caption{Proton factorial cumulants ($\kappa_5$, $\kappa_6$) and net-proton cumulant ratios ($C_{5}/C_{1}$, $C_{6}/C_{2}$) in Au+Au collisions at $\sqrt{s_{NN}} = 7.7$--27~GeV as a function of collision centrality, quantified by the average number of participant nucleons $\langle N_{\mathrm{part}} \rangle$. BES-II results are compared with those from BES-I where applicable. Also shown is a comparison between BES-II measurements using two different centrality estimators: RefMult3X (red) and RefMult3 (blue); see text for details. The RefMult3X data points (red) are slightly shifted along the positive x axis for better visualization. Error bars and bands represent statistical and systematic uncertainties, respectively. For clarity in presentation, BES-I data points for 0-40\% centrality (both factorial cumulants and cumulant ratios) are scaled down by a factor of 10 at each beam energy. Similarly, for the 0-40\% centrality class in BES-II, the factorial cumulant data ($\kappa_5$, $\kappa_6$) from RefMult3 and RefMult3X are scaled down by a factor of 10 uniformly across all energies. No scaling is applied to the BES-II net-proton cumulant ratio data ($C_{5}/C_{1}$, $C_{6}/C_{2}$).}
  \label{label_fig2}
\end{figure*}

Figure~\ref{label_fig2} shows the centrality ($\langle N_{\mathrm{part}} \rangle$: average number of participating nucleons) dependence of the fifth- and sixth-order proton factorial cumulants ($\kappa_5$, $\kappa_6$) and the net-proton cumulant ratios ($C_5/C_1$, $C_6/C_2$). Results are presented for 0--40\%, 40--50\%, 50--60\%, 60--70\%, and 70--80\% centrality bins. The typical $\langle N_{\mathrm{part}} \rangle$ values corresponding to these centrality bins can be found in Ref.~\cite{npart_cent}.
The broader 0--40\% bin is used to improve statistical precision. Overall, a weak centrality dependence is observed.

For the 0--40\% bin, measurements using RefMult3 and RefMult3X centrality definitions agree within $0.4\sigma$ of each other, indicating that centrality resolution effects are minimal in this range. In more peripheral bins, slightly lower net-proton \(C_{5}/C_{1}\) values are seen with RefMult3X, reflecting centrality resolution effects. Similar trends were observed in previously reported fourth-order proton fluctuation results from BES-II~\cite{STAR:2025zdq}. The current fifth- and sixth-order measurements using RefMult3 show good consistency with BES-I results, with differences remaining within $1.7\sigma$ of BES-I. Notably, the BES-II data exhibit significantly reduced uncertainties. For example, at $\sqrt{s_{NN}} = 19.6$~GeV in 0--40\% centrality, the total uncertainty on $C_6/C_2$ is reduced by a factor of 6 compared to BES-I.

The energy dependence of these observables is explored in Fig.~\ref{label_fig3}, which shows $\kappa_4$, $\kappa_5$, and $\kappa_6$ as functions of $\sqrt{s_{NN}}$ for 0--40\% and 70--80\% centralities. Within uncertainties, the 0–40\% results exhibit a largely flat energy dependence with values fluctuating around zero, while the 70–80\% data remain closest to the Poisson baseline (zero).

Figure~\ref{label_fig4} presents the net-proton cumulant ratios. A mild energy dependence is observed in $C_4/C_2$, with values decreasing down to around 20~GeV and showing a possible rise at lower energies. In contrast, $C_5/C_1$ and $C_6/C_2$ display little or no significant variation with beam energy.
A suppression of $C_4/C_2$ relative to the Poisson expectation (unity) is observed across the measured energies, with a more pronounced deviation in the 0–40\% centrality bin compared to 70–80\%.  These trends remain well below the values reported by STAR at  $\sqrt{s_{NN}}$ = 3~GeV, which were obtained within the narrower rapidity range of $-0.5 < y < 0$~\cite{STAR:2022vlo}.

\section{Model Comparisons}

To interpret the measured cumulants, we compare them with a range of theoretical and phenomenological models representing different QCD regimes.

Lattice QCD provides first-principles predictions for cumulant ratios at small baryon chemical potential, using Taylor expansions in $\mu_B/T$~\cite{Bazavov:2020bjn}. These calculations predict a specific ordering of cumulants, namely $C_3/C_1 > C_4/C_2 > C_5/C_1 > C_6/C_2$, consistent with expectations from a smooth crossover. FRG calculations extend these predictions to larger $\mu_B$ values and incorporate critical fluctuations. They forecast stronger deviations in cumulant ratios near a critical point~\cite{Fu:2021oaw}. Caution is warranted when comparing with lattice QCD and FRG results, which compute net-baryon fluctuations in a grand canonical ensemble and do not account for experimental acceptance. In contrast, experiments measure net-proton number fluctuations —used as a proxy for net-baryon number— within finite experimental acceptances. Hadron Resonance Gas with Canonical Ensemble treatment (HRG-CE) assumes a thermalized hadron gas with exact baryon number conservation~\cite{Braun-Munzinger:2020jbk}. UrQMD is a hadronic transport model that does not include a phase transition. It serves as a baseline for non-critical behavior driven by hadronic interactions and resonance decays~\cite{Bass:1998ca,Bleicher:1999xi}. A two-component model is a simplified analytical approach inspired by a first-order phase transition. It assumes contributions from two distinct event classes—hadronic and mixed phases—resulting in a bimodal multiplicity distribution. This structure can lead to sign alternation and rising magnitudes in factorial cumulants~\cite{Bzdak:2018axe}.

\begin{figure*}[!htbp]
\centering
  \includegraphics[scale=0.95]{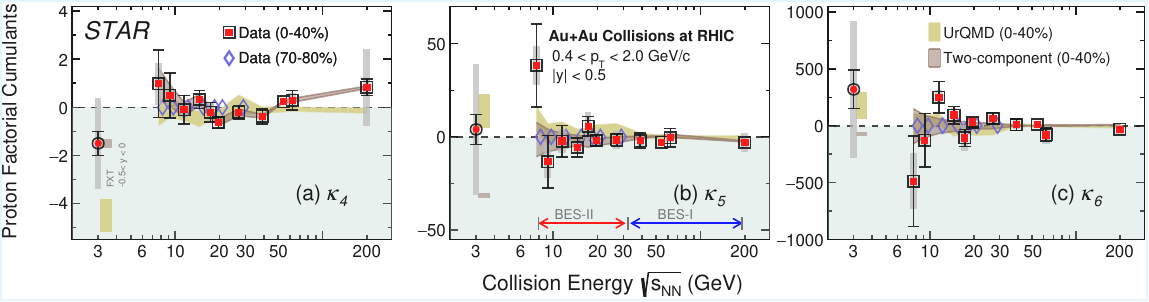}
  \caption{Proton factorial cumulants ($\kappa_4$, $\kappa_5$, $\kappa_6$) in Au+Au collisions at $\sqrt{s_{NN}} = 7.7$--27~GeV (BES-II) and $\sqrt{s_{NN}} = 39$--200~GeV (BES-I)~\cite{STAR:2022vlo, STAR:2025zdq}, shown as a function of collision energy. Results at $\sqrt{s_{NN}} = 3$~GeV from the fixed-target (FXT) program~\cite{STAR:2022vlo} are also shown. Error bars and shaded bands represent statistical and systematic uncertainties, respectively. Model calculations from UrQMD and the two-component model are shown for comparison.}
  \label{label_fig3}
\end{figure*}

Figure~\ref{label_fig3} compares the measured proton factorial cumulants with the two-component model expectations. Following Ref.~\cite{Bzdak:2018axe}, the model uses a combination of Poisson and Binomial distributions, with parameters constrained by the measured values of $\kappa_n$ up to fourth order. For 0–40\% centrality, both the data and the model show no sign change within uncertainties—an expected feature near a first-order phase transition—while the central values of the data hint at an increase with order, but with marginal significance. These trends are also qualitatively reproduced by the UrQMD model. In contrast, the 70--80\% peripheral data remain close to zero across all energies.

\begin{figure*}[!htbp]
\centering
  \includegraphics[scale=0.95]{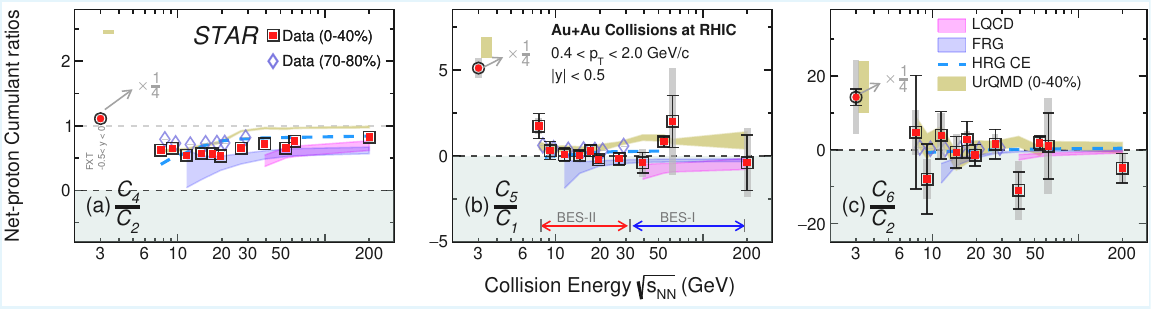}
  \caption{Net-proton cumulant ratios ($C_4/C_2$, $C_5/C_1$, $C_6/C_2$) in Au+Au collisions at $\sqrt{s_{NN}} = 7.7$--27~GeV (BES-II) and $\sqrt{s_{NN}} = 39$--200~GeV (BES-I)~\cite{STAR:2020tga, STAR:2022vlo, STAR:2025zdq}, shown as a function of collision energy. Results at $\sqrt{s_{NN}} = 3$~GeV from the fixed-target (FXT) program~\cite{STAR:2022vlo} are also shown. Error bars and shaded bands indicate statistical and systematic uncertainties, respectively. Model calculations from lattice QCD, FRG, HRG (canonical ensemble), and UrQMD are included for comparison. At 3\, GeV, cumulant ratios (filled red circles), $C_{4}/C_{2}$, $C_{5}/C_{1}$, and $C_{6}/C_{2}$, are scaled down by a factor of 4 to visually accommodate other data points.
}
  \label{label_fig4}
\end{figure*}

\begin{figure*}[!htbp]
\centering
  \includegraphics[scale=0.75]{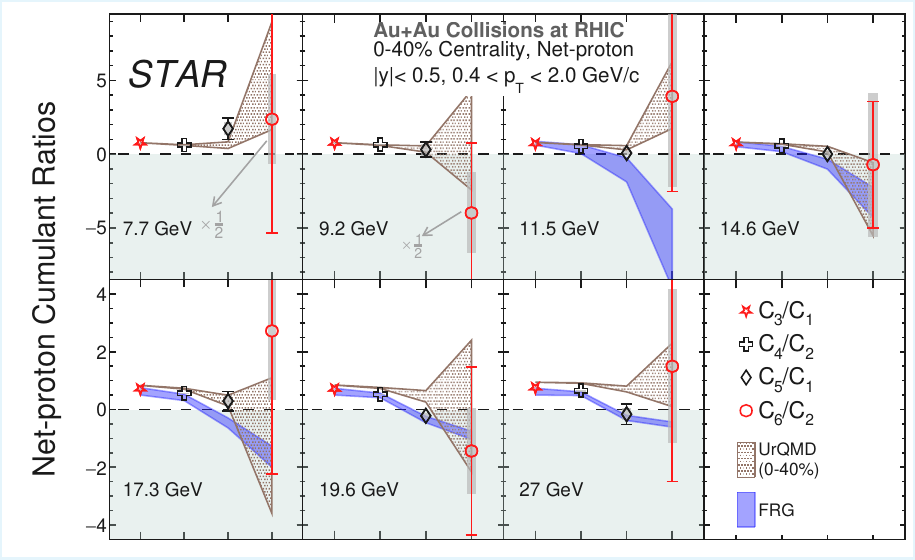}
  \caption{Cumulant ratios $C_3/C_1$ (stars), $C_4/C_2$ (crosses), $C_5/C_1$ (diamonds), and $C_6/C_2$ (circles) for net-proton distributions in 0--40\% centrality Au+Au collisions at $\sqrt{s_{NN}} = 7.7$--27~GeV (BES-II). The shaded bands represent predictions from FRG and UrQMD models. At 7.7 and 9.2 GeV, the $C_6/C_2$ points (open red circles) are scaled down  by a factor of 2 for clarity of presentation.}
  \label{label_fig5}
\end{figure*}

Figure~\ref{label_fig4} shows the energy dependence of the net-proton cumulant ratios and compares them with various theoretical predictions. In 0--40\% centrality Au+Au collisions, the $C_5/C_1$ and $C_6/C_2$ ratios fluctuate around zero. The current measurements of $C_{6}/C_{2}$ yield a significance of about $0.2\sigma$ for observing negative values at more than half of the collision energies in the range $\sqrt{s_{NN}}=7.7$--$200~\mathrm{GeV}$, compared to $1.7\sigma$ obtained in BES-I~\cite{STAR:2022vlo}. Within uncertainties, these cumulant ratios are consistent with the values predicted by lattice QCD and FRG for a smooth crossover, as well as with those expected from models without a critical point such as UrQMD and HRG-CE. In the peripheral 70--80\% centrality bin and in STAR’s fixed-target measurement at $\sqrt{s_{NN}} = 3$~GeV~\cite{STAR:2022vlo}, the ratios remain positive. The $C_4/C_2$ ratio is also positive at all energies.

Overall, the behavior of $C_4/C_2$ and $C_5/C_1$ in 0--40\% collisions is well described by lattice QCD, FRG, and HRG-CE at $\sqrt{s_{NN}} \gtrsim 27$~GeV, whereas UrQMD better reproduces the measurements at lower energies ($\lesssim 11.5$~GeV).


The cumulant ratios are also used to test the predicted ordering $C_3/C_1 > C_4/C_2 > C_5/C_1 > C_6/C_2$, which is expected for a smooth crossover scenario in lattice QCD and FRG frameworks. A notable violation of this ordering has been reported only in STAR’s fixed-target data at $\sqrt{s_{NN}} = 3$~GeV~\cite{STAR:2022vlo}. As shown in Fig.~\ref{label_fig5}, the BES-II data in 0--40\% centrality follow the expected hierarchy in the energy range $\sqrt{s_{NN}} = 9.2$--27~GeV, although uncertainties in $C_6/C_2$ remain large. At 7.7~GeV, a clear ordering trend is not evident. The probability of the cumulant ratios up to fifth order following the predicted ordering, assessed via a statistical test as done in Ref.~\cite{STAR:2022vlo}, ranges from 63.8\% to 99.5\% over $\sqrt{s_{NN}} = 9.2$--27 GeV and decreases substantially to 8.5\% at 7.7 GeV. While UrQMD calculations also seem to follow the expected ordering, the $C_5/C_1$ values remain consistently positive, deviating more noticeably from the data at higher energies.

\section{Summary}

We report new high-statistics measurements of the fifth- and sixth-order factorial cumulants ($\kappa_5$, $\kappa_6$) and cumulant ratios ($C_5/C_1$, $C_6/C_2$) of (net-)proton multiplicity distributions in Au+Au collisions at $\sqrt{s_{NN}} = 7.7$--27~GeV, corresponding to a baryon chemical potential ($\mu_B$) range at chemical freeze-out of approximately 400--150~MeV~\cite{STAR:2017sal}. While consistent with earlier BES-I results, these BES-II measurements offer significantly improved precision. Notably, the total uncertainty in $C_6/C_2$ at 19.6 GeV for 0–40\% centrality is reduced by a factor of  6. The higher-order cumulants show a weak dependence on both collision centrality and energy.

The higher-order proton factorial cumulants $\kappa_4$, $\kappa_5$, and $\kappa_6$ increase in magnitude with order but show no sign alternation within uncertainties in collisions at 0-40\% centrality across BES-II energies. This suggests that there is no evidence for a two-component structure that could signal a first-order phase transition. In peripheral (70–80\%) collisions, all higher-order factorial cumulants remain consistent with zero.

The net-proton cumulant ratios $C_5/C_1$ and $C_6/C_2$ fluctuate around zero in collisions at 0-40\% centrality, and their signs—within uncertainties—are consistent both with lattice-QCD predictions for a smooth crossover and with expectations from UrQMD, which has no phase transition. At higher energies ($\sqrt{s_{NN}} \gtrsim 27$ GeV), the trends in $C_4/C_2$ and $C_5/C_1$ agree with lattice QCD, FRG, and HRG-CE calculations, while UrQMD provides a better description at lower energies ($\lesssim 11.5$~GeV).

Moreover, the measured cumulant ratios follow the QCD-thermodynamic ordering 
$C_3/C_1 > C_4/C_2 > C_5/C_1 > C_6/C_2$ in the range 
$\sqrt{s_{NN}} = 9.2$--27~GeV, although the uncertainties in $C_6/C_2$ are still large. 
No clear ordering is observed at 7.7~GeV. 
UrQMD shows a similar trend, with $C_5/C_1$ remaining positive and deviating from the data at higher energies. Overall, these results offer important insights into baryon number fluctuations and provide valuable input for mapping the QCD phase structure under extreme conditions.

We thank the RHIC Operations Group and SDCC at BNL, the NERSC Center at LBNL, and the Open Science Grid consortium for providing resources and support.  This work was supported in part by the Office of Nuclear Physics within the U.S. DOE Office of Science, the U.S. National Science Foundation, National Natural Science Foundation of China, Chinese Academy of Science, the Ministry of Science and Technology of China and the Chinese Ministry of Education, NSTC Taipei, the National Research Foundation of Korea, Czech Science Foundation and Ministry of Education, Youth and Sports of the Czech Republic, Hungarian National Research, Development and Innovation Office, New National Excellency Programme of the Hungarian Ministry of Human Capacities, Department of Atomic Energy and Department of Science and Technology of the Government of India, the National Science Centre and WUT ID-UB of Poland, the Ministry of Science, Education and Sports of the Republic of Croatia, German Bundesministerium f\"ur Bildung, Wissenschaft, Forschung and Technologie (BMBF), Helmholtz Association, Ministry of Education, Culture, Sports, Science, and Technology (MEXT), and Japan Society for the Promotion of Science (JSPS).

\bibliographystyle{apsrev4-1} 
\bibliography{ReferencesC6_list} 

\end{document}